\documentclass[10pt,conference,USletter]{IEEEtran}
\IEEEoverridecommandlockouts
\usepackage{cite}
\usepackage{amsmath,amssymb,amsfonts}
\usepackage{algorithmic}
\usepackage{fancyhdr}
\usepackage{graphicx}
\usepackage{siunitx}
\usepackage{textcomp}
\usepackage{xcolor}
\usepackage{hyperref}
\usepackage[singlespacing]{setspace}
\renewcommand\IEEEkeywordsname{Keywords}
\def\BibTeX{{\rm B\kern-.05em{\sc i\kern-.025em b}\kern-.08em
    T\kern-.1667em\lower.7ex\hbox{E}\kern-.125emX}}

\begin{document}

\title{Mobile Edge Computing, Metaverse, 6G Wireless Communications,  Artificial Intelligence, and Blockchain: Survey and Their Convergence}

\author{\IEEEauthorblockN{Yitong Wang}
\IEEEauthorblockA{\textit{School of Computer Science and Engineering} \\
\textit{Nanyang Technological University}\\
Singapore\\yitong.wang@ntu.edu.sg \\
}
\and
\IEEEauthorblockN{Jun Zhao}
\IEEEauthorblockA{\textit{School of Computer Science and Engineering} \\
\textit{Nanyang Technological University}\\
Singapore\\ junzhao@ntu.edu.sg 
}
}

\maketitle
 \thispagestyle{fancy}
\pagestyle{fancy}
\lhead{This paper appears in the Proceedings of 2022 IEEE 8th World Forum on Internet of Things (WF-IoT).\\ Please feel free to contact us for questions or remarks.}

\cfoot{~\\[-25pt]\thepage}

\begin{abstract}
With the advances of the Internet of Things (IoT) and 5G/6G wireless communications, the paradigms of mobile computing have developed dramatically in recent years, from centralized mobile cloud computing to distributed fog computing and mobile edge computing (MEC). MEC pushes compute-intensive assignments to the edge of the network and brings resources as close to the endpoints as possible, addressing the shortcomings of mobile devices with regard to storage space, resource optimisation, computational performance and efficiency. Compared to cloud computing, as the distributed and closer infrastructure, the convergence of MEC with other emerging technologies, including the Metaverse, 6G wireless communications, artificial intelligence (AI), and blockchain, also solves the problems of network resource allocation, more network load as well as latency requirements. Accordingly, this paper investigates the computational paradigms used to meet the stringent requirements of modern applications. The application scenarios of MEC in mobile augmented reality (MAR) are provided. Furthermore, this survey presents the motivation of MEC-based Metaverse and introduces the applications of MEC to the Metaverse. Particular emphasis is given on a set of technical fusions mentioned above, e.g., 6G with MEC paradigm, MEC strengthened by blockchain, etc.
\end{abstract}

\section{Introduction}
The exponential increase in data traffic, the number of devices and the variety of service scenarios are the reasons for the advent of 5G. In the 5G era, mobile networks will serve not only mobile phones but also various types of devices such as tablets, mobile vehicles, and various sensors. The service scenarios are thus becoming increasingly diverse, such as mobile broadband, mission-critical internet, and large-scale machine-type communications. The inherent attributes of 5G mobile networks include enhanced mobile broadband (eMBB),  ultra-reliable low-latency communications (URLLC), and massive machine type communications (mMTC), driving the implementation of a broad range of applications \cite{you2021towards}.

To meet the requirements of the fast growing mobile world with high data rates and low latency, 5G networks have proposed the following evolutionary goals: network functions virtualisation (NFV), software defined networking (SDN) and user-plane functionality of the core network closer to the base station. In this context, MEC is considered a fundamental technology for the transition to 5G, as it enables a more brilliant mobile network by combining the Internet service environment on the network edge with cloud computing. MEC emphasizes proximity to the user, reduces network operation and service delivery latency, and improves user experience. Many commercial operators and academic experts are currently working on MEC to enable mobile network operators to expose more network control capabilities to third party developers.

The European Telecommunications Standards Institute (ETSI) defines MEC as providing an IT service infrastructure in close proximity to the subscribers and the equivalent computing power of cloud servers \cite{computing2014mobile}. The MEC can be thought of as a cloud server deployed at nodes close to the clients to handle tasks that cannot be tackled by the traditional infrastructure, such as machine-to-machine (M2M) gateways, control functions, intelligent video acceleration, etc. It is important to note that the ``M" in MEC initially stood for ``mobile", referring to the mobile network environment. Later ETSI  extended the definition of ``M" to ``multi-access", with the aim of extending the concept of edge computing to non-3GPP access scenarios. The term ``mobile edge computing" is also gradually being transitioned to ``multi-access edge computing (MEC)".

MEC runs at the edge of the network and is convenient for security-critical applications. In addition, MEC servers usually have a high level of computing power, making them particularly suitable for analyzing and processing large amounts of data. At the same time, the geographical proximity of the MEC to the user or information source reduces the latency of the network to answer user requests and reduces the likelihood of network congestion in the transport and core parts of the network. Finally, MEC servers can obtain real-time network data, allowing for link-aware adaptation and offering the possibility of deploying location-based applications, which can significantly improve the service quality. 

The sixth generation of mobile communication systems (6G) meet the requirements of a comprehensive connected and smart digital environment \cite{chang20226g,zhao2021survey}, supporting complete network coverage, enhanced intelligent applications and stronger security. Compared to previous generations of mobile communication technology, 6G will have a radical change from ``connected things" to ``connected intelligence", i.e., going beyond personalized communication to realize intelligent connections between people, devices, and resources. In the 6G era, interaction of various information will be further developed from augmented reality/virtual reality (AR/VR) to high-fidelity extended/mixed reality (XR/MR) interaction and even holographic communication. This vital shift relies on XR fully mobilizing the senses of sight, touch, hearing, and smell to make customers enjoy fully immersive holographic experiences such as virtual sports, virtual travel, and virtual games anytime and anywhere. Based on 5G autopilot, the 6G era will see smarter and more diverse transportation used by humans, enabling on-demand, point-to-point smart travel. 

Artificial intelligence (AI) algorithms help machines learn from vast amounts of data, constantly adapting to various environments and exhibiting human-like behaviours to help people carry out daily activities. Traditionally, AI technologies consist of supervised learning, unsupervised learning as well as reinforcement learning. AI technology was in the limelight when AlphaGo beat the world's Go players by learning data from 300,000 games in 2016\cite{yang2019federated}. In the 6G system, AI technology can be used to solve diverse tasks with advanced algorithms, for example, allocation of computing resources, defence against attacks, smart decisions under uncertain environments, etc. At the same time, personalised services can be realised through training and analysis of user data \cite{pham2022artificial}.

Blockchain, as a decentralised technology and an economic system, enables not only the protection of users' assets and data but also allows for autonomous user control, creating a user-centric network. Blockchain can be divided into three kinds, including public, consortium, and private, adapting to different scenarios. Importantly, this decentralised network will avoid some critical risks, for example, low transparency and Single Point of Failure (SPoF).

\subsection{Recent Work on Mobile Edge Computing}
Current research on MEC tends to focus on the design of computation offloading and energy consumption strategy. To tackle the coupling challenge between energy consumption and finer-grained offloading, the algorithm proposed by Guo \textit{et al.} \cite{guo2021lyapunov} is effective in reducing latency and meeting the energy consumption of multiple users by building models for the energy harvesting process. Wang~\textit{et~al.}~\cite{wang2020game} present dynamic task offloading strategies applied to vehicular networks. An optimal response mechanism is investigated to improve the vehicle system's efficiency in utilising the resources of MEC systems, taking into account the vehicle's route and the resource requirements of different vehicles.

\subsection{Contribution of the Paper}
As the demand for data process is becoming more strict, MEC is gaining widespread use and high acclaim. This paper provides a systematic overview of MEC technology, firstly introducing the basic concepts of different computing paradigms and detailing the reference framework of each computing. This is followed by an introduction to the key technologies and application scenarios of future MEC technologies. Then the challenges that MEC will address in existing and future networks are presented, as well as a summary of several existing problems in MEC that have yet to be solved. Notably, the integration of the emerging metaverse system and MEC is presented, along with the associated motivations and challenges. Last but not least, the technical convergence of MEC with 6G, AI,  blockchain and mateverse technologies, and potential research directions are discussed. 

\subsection{Paper Organization}
The paper is structured in the following way. Section~\ref{sectionII} provides an overview of the various computing paradigms and related technologies that have evolved over the last decade, along with an analysis of their strengths and weaknesses. Next, Section~\ref{sectionIII} presents an overview of recent applications of MEC in different scenarios, with a particular focus on Mobile Augmented Reality (MAR). Section~\ref{sectionIV} presents   future open challenges of MEC. The motivation and future vision of  MEC-based Metaverse are introduced in Section~\ref{sectionV}. Significantly, Section~\ref{sectionVI} investigates the convergence of MEC-based Metaverse with 6G wireless communications, AI, and blockchain. Lastly, we summarise the paper and offer a vision for the work ahead in Section~\ref{sectionVII}.

\section{Modern Computing Paradigms}\label{sectionII}
The remarkable expansion of mobile internet traffic flows over the past decade has driven a sharp evolution in computing paradigms. Especially cloud computing, as the dominant computing option, can fulfil most of the users' usage requirements. Nevertheless, with the increasing advancement of wireless communication technologies, modern mobile applications have more stringent requirements that conventional cloud computing cannot cope with anymore, such as ultra-low latency,  ultra-high reliability and data rate, and high spectral efficiency. Meanwhile, big data-associated Internet of Things devices cause a large growth in the total traffic flows\cite{linthicum2017connecting}. To break the limitations of cloud computing and ensure better customer service performance, emerging computing paradigms, especially mobile edge computing, have gained the spotlight in both commercial and academic fields. This section illustrates the various potential computing architectures currently available in wireless communication systems, covering cloud computing, fog computing, and edge computing. Furthermore, the benefits and drawbacks of mobile edge computing architecture are deeply investigated.

\subsection{Cloud Computing}
The emergence of cloud computing has drastically modified the mode of communication between servers and end users. According to NIST definition, ``Cloud computing is a model that enables shared network access to various configurable system resources, including computing resources, storage space, applications, etc., which can be quickly reconfigured and deployed with minimal cost and the involvement of service providers” \cite{mell2011nist}. Three fundamental services are supplied by cloud computing to users: Platform as a Service (PaaS), Software as a Service (SaaS),  as well as Infrastructure as a Service (IaaS). The deployment of this paradigm consists of four patterns, covering private cloud computing, community cloud computing, public cloud computing, and hybrid cloud computing.

\subsection{Mobile Cloud Computing}
Phenomenal developments in cloud computing have reduced expenditure on various resources and separated the underlying infrastructure from the services, enabling users to achieve on-demand self-service. Smart mobile devices (SMDs) have also gained tremendous popularity due to the great support they receive from consumers. With SMDs, users can enjoy a variety of online web services such as gaming, video, conferencing, shopping and more. However, as the complexity of SMDs grows and the demands on computing resources get more stringent, various mobile applications are placing greater demands on the memory and battery life of the corresponding devices. Current smartphones are not sufficient to cope with these requirements. Consequently, given these constraints, users are still unable to enjoy some of the applications on their smart mobile devices.

According to \cite{othman2013survey}, mobile cloud computing (MCC) combines cloud computing technologies with mobile devices to meet the stringent requirements of mobile equipment concerning computing resources, storage, energy, and context-aware capabilities. MCC is an emerging model of distributed computing that extends computing to smart mobile devices. Typically, to overcome these limitations, a lightweight cloud server known as Cloudlet is deployed at the network's edge sides.

Computation offloading is an integral process in MCC, offloading resource-intensive computation and energy-intensive applications from end devices to resource-rich cloud servers for resource-saving purposes. However, traditional computation offloading tasks are not concerned about bandwidth and power consumption, and cannot be used directly on smartphones. Hence, it is crucial to design an application model that supports offloading. There are currently many papers on computation offloading. For example, Khan \textit{et al.} \cite{othman2013survey} summarize and analyze some latest MCC application models, further pointing to future research directions. Li \textit{et al.} \cite{zhang2016survey} review and compare existing computation offloading models.

\subsection{Fog Computing}
The development of emerging technologies like the Internet of Things (IoT) requires low-latency computing to complete the applications implemented. Due to the on-demand service and scalability features of cloud computing, cloud infrastructures can process data generated by IoT devices. However, cloud processing requires a lot of bandwidth and storage, so processing only through the cloud is not the best option. As IoT environments and other time-sensitive applications have strict constraints on latency, fog computing technologies are rapidly evolving.

The term ``Fog Computing" was coined by researchers at Cisco Systems in 2012 \cite{bonomi2012fog}. The concept of fog computing is not the same as cloud computing, and is a distributed computing concept. Mostly, fog computing and cloud infrastructure are independent of each other. Fog computing is located between the cloud servers and IoT devices, and this computing paradigm uses devices with computing power for data processing, e.g., routers, base stations, smartphones with multiple cores, etc. These devices can act as fog devices. Fog computing also utilizes free computing resources near the customers to provide more efficient and faster services. These devices are involved in computing at peak times but idle at other times. At the same time, fog computing can still rely on cloud servers when it comes to complex computations. 

\subsection{Edge Computing}
Fog computing breaks many of the limitations that exist in cloud computing and improves the user experience. However, edge computing has been developed to further meet the requirements of ultra-low latency, stability, and reliability. Compared to fog computing, edge computing has lower latency, and the end devices are not identical. In edge computing, there are three distinct types of devices: sensors, end-user devices (e.g., smartphones), and servers. The raw data is generated in the sensors, received by the end-user devices, and then transmitted to the servers for processing. In fog computing, however, only the servers are included. 

\textbf{Personal Edge.} This type of edge computing is centred on the individual user and exists around the user, such as a typical user's residential environment. Also, this type of computing is applied to the users' various intelligent devices, for example, home robots, smart glasses, smartphones, smartwatches and home automation systems. The most characteristic feature of the personal edge is that it is not stationary. Personal edge computing devices remain mobile as consumers move around in various scenarios. As the demand for applications grows and the variety of personal devices increases, the personal edge has enormous growth potential.

\textbf{Business Edge.} The business edge is the most sought-after category of edge computing. The reasons are as follows: 1) Numerous application scenarios. Whether in the office or a place of entertainment, there is a need for commercial edge computing to process a wide variety of tasks. 2) ``Sense-Process-Act''. The unique way of processing missions has led to the commercial edge growing more and more rapidly in the Industrial IoT and is considered the unique edge.

\textbf{Cloudy Edge.} This is the most rarely discussed type of edge computing, but it is the earliest form of edge. Cloudy edge is a topological word that refers to the edge of a service supplier or business environment network. The services go from a dial-up modem to a residential or remote office in the first instance.

\subsection{Mobile Edge Computing}
Mobile edge computing has become a pivotal technology as latency and energy requirements for various mobile applications increase. As stated by the European Telecommunications Standards Institute (ETSI) in 2014, mobile edge computing (MEC) is described as a proposed novel framework ``to deliver IT and cloud computing functionality inside the radio access network (RAN), near the mobile customers" \cite{computing2014mobile}. In general, there are fields of overlap between fog computing and mobile edge computing, and some terminologies are interchangeable in specific contexts. MEC draws on the concept of cloud computing, where new devices called MEC servers are deployed at the network edge to provide more computing resources and storage space in order to reduce latency and provide context-aware services to enhance the service performance. With this computing model, users can also enjoy caching services. The MEC servers can deliver locally cached content directly to the users, thus reducing transmission latency and the need for users to communicate with the cloud servers.

\textit{1)	Pros of Mobile Edge Computing:}

\textit{Latency:} The MEC servers are deployed in base stations (BSs), which means that the distance over which information is transferred is much smaller than in cloud computing, thus effectively reducing transmission latency. At the same time, the computing power of the BSs is gradually increasing, and although it is not on par with the cloud servers which can process data from many users simultaneously,  the latency of the MEC is also much reduced. Overall, MEC is effective in reducing End-to-End (E2E) latency. 
\textit{Quality of Service/Quality of Experience (QoS/QoE):} When users request services, the content cached in the edge servers can be transmitted to the end devices, which ensures that the content is error-free and reduces latency, so QoS/QoE is greatly guaranteed. 
\textit{Energy:} A pivotal feature of MEC is its support for computation offloading so that computation assignments could be offloaded to the MEC servers, reducing any involvement of mobile devices. This will, therefore, significantly increase the battery life of smart devices, resulting in significant and influential energy savings. 
\textit{Context-awareness:} The MEC servers can get almost as close as possible to the users' terminal devices, grasp the users' behaviours and location information in real time and provide the appropriate personalized services by analyzing the users' surroundings.

\textit{2)	Cons of Mobile Edge Computing:}
\textit{Operation cost:} Compared to centralized cloud computing, MEC requires a higher operation cost to perform updates when the servers need to be renewed.

\section{Mobile Edge Computing Application Scenarios}\label{sectionIII}
Application scenarios for MEC can be categorised in three main groups: local triage, data services, and service optimization. Local triage is mainly applied to transmission-constrained and latency-reducing scenarios, including enterprise campuses, campuses, popular venues, local video surveillance, VR/AR scenarios, local live video streaming, and edge content distribution networks (CDN). Data services include indoor positioning, Telematics, etc. Service optimization includes video quality of service (QoS) optimization, live video streaming and game acceleration, etc. By introducing intelligent computing capabilities on the base station side, the challenges mentioned above for operators and network service providers are solved, while wireless resources are managed more intelligently and optimally, and different levels of services can be realized. In this section, we focus on the application of MEC in mobile augmented reality.

According to paper \cite{siriwardhana2021survey}, Mobile Augmented Reality (MAR) applications enable users to access AR services through mobile interfaces, by running AR applications on the device sides. With the development of MEC technology, MAR applications are guaranteed to be implemented. We will present the following areas.

\textbf{Smart City/Home.} At the Smart City level, MAR with MEC helps the public to identify and visualize building structures. This technology can also help people to enjoy better navigation service when travelling. In the smart home, the technology allows occupants to control switches and operate appliances remotely. As an example, IKEA is a typical example of a MAR service, which allows customers to select furniture and appliances in a shopping mall by using the MAR application on their mobile devices to see how various pieces of furniture will look in their homes and to simulate the colour and style of the furniture, thus helping them to select the furniture to their satisfaction. When buying clothes, customers can also use the app to dress virtually and see how their clothes look. All these applications make people's daily lives more convenient. By deploying MAR servers to the MEC, smart cities and homes will take an even more significant leap forward.

\textbf{Vehicles.} By applying MAR with MEC to vehicle driving systems, navigation information, road conditions and surrounding vehicle information can be projected onto the vehicle windscreen in real time, enabling drivers to grasp traffic information and make accurate judgements without distraction quickly. At the same time, the deployment of the MEC server allows all vehicles in the current service area to receive the information shared by other vehicles. For example, when there are other vehicles in front of the drivers blocking the view of the road ahead so they can avoid risks in time. The AR view can also be used for reference when manufacturing, inspecting and assembling vehicle components, comparing the real components with the projection, and assembling them according to the projection, increasing production efficiency. 

\section{Future with Mobile Edge Computing}\label{sectionIV}
\subsection{Offloading Decisions}
Presently, there are a variety of offloading policies, each with its own advantages and disadvantages, apart from cloud computing, the other policies are listed as follows:

\textbf{Full Computing.} 1) Local Computing: All computing tasks are processed in the local user equipment (UE) and not offloaded to the MEC servers so that the latency generated by local execution is much smaller than that of offloading computing assignments to the servers. However, this approach also places certain demands on the battery life and energy consumption of the UE. This will also place particular demands on the end devices' computation power. 2) Edge Computing: Unlike the concept of local computing, edge computing offloads all compute-intensive tasks to the MEC servers for processing. Then there are three types of latency to be concerned about: a) the time to offload the data to the server; b) the duration required for the MEC servers to process the data; and c) the period required for the UE to receive the processed data. We expect that the sum of these three delays is smaller than the local computing latency. In this way, the UE energy is also saved.

\textbf{Partial Computing.} For partial computing, we split the processing mobile applications into two parts: the non-offloaded and offloaded parts, which is more flexible than the binary decision of local computing or edge computing. The \mbox{non-offloaded}   parts will be computed on the local UE, and the offloaded parts will be sent to the edge side for processing. However, accurately separating the two components and balancing the constraints are challenging.
In addition, offloading determinations are achieved by taking into account various elements, such as energy consumption, latency and network bandwidth. Wu \textit{et al.} \cite{wu2021deep} investigate a new joint optimization algorithm to optimise the system cost with the constraints of network bandwidth and system computation resources   in 5G-based multi-access edge computing network for vehicle awareness. Yang \textit{et al.} \cite{yang2022intelligent} supply design a deep reinforcement learning approach to the computing offloading in a MEC-based cooperative vehicle infrastructure system. They jointly take the accuracy, energy consumption and execution latency of the system into consideration. 

The offloading decision is an essential part of the MEC, as it determines whether the computation needs to be full computing or partial computing. When making the decision, we need to take into account not only the energy consumption on the UE side but also the energy spent by the MEC servers. It is also worth considering the case of a single device offloading computation to multiple servers (e.g., for very stringent delay requirements) and the case of multiple components offloading. Crucially, most studies assume that the UE is stationary when making decisions. In reality, UEs may be on the move, and there are situations where they move from one cluster to another. In this case, the decision to minimize the cost subject to the mobility of UEs is a concern.

\subsection{Mobility Management}
In MEC systems, most MEC applications rely heavily on the states and previous behaviours of the users. Therefore, mobility is a key consideration when implementing these applications. Through user mobility and excellent mobility management, we can ensure application mobility and seamless service to enhance the user experience and QoS further. Consequently, how to implement mobility management is an essential concern for researchers. Currently, there are several ways to address mobility management, as explained below:

\textbf{Caching.} As data traffic increases exponentially (e.g., VR video), the variety of data processing can saturate the network capacity. The Motion-To-Photon (MTP) latency for VR video viewing should be no greater than 20ms \cite{ohl2015latency}. Caching is defined as storing popular or repeatedly served contents in advance to the base servers at the edge of the network during off-peak hours, which can effectively reduce the high latency and poor QoS associated with mobility and provide stable services to users. Computations related to visual images rendering can be offloaded from the UE to the MEC servers. The total latency is further reduced by the processing of the MEC servers.

As we can appreciate from the paper \cite{liu2021rendering}, the variables that need to be considered in caching are: how can the popularity of the content to be cached be accurately predicted? How can we achieve collaborative caching across multiple cells to improve web caching performance? The solution proposed by Liu~\textit{et~al.}~\cite{liu2021rendering} is 1) to slice the video content to provide a multi-cell placement and sharing scheme, 2) to predict the popularity level of each video by the proposed prediction method, and 3) to deploy the rendering and caching operations of the videos together on the MEC servers to further enhance the network characteristics by utilizing the high computing power of the servers. As a result, when the cached content is shared among MEC servers in multiple cells, the stability of the content and the continuity of the service is ensured when users are on the move.

\textbf{SDN-Controller.} Conventional IP networks have numerous drawbacks, including configuration complexity and new network features that are extremely difficult to be developed and deployed. SDN is characterised by splitting the control layer from the data layer, with the control logic being transformed into the SDN controller. The application of the SDN concept to the MEC system not only provides a unified control plane interface and ensures the reliability of services but also effectively manages the heterogeneity of the various service requests.

Shah \textit{et al.} \cite{shah2022sdn} propose a SDN-based MEC-enabled 5G vehicular networks. They use four software modules developed by themselves in the SDN controller to further design the rules of QoS. Furthermore, the SDN controller proposed in the paper can federate and coordinate the allocation of MEC resources in order to further provide continuous service and seamless coverage to mobile users.

\section{Mobile Edge Computing for the Metaverse}\label{sectionV}

In this section, we discuss the applications of MEC to the Metaverse. Section~\ref{sec-Metaverse} provides a brief overview of the Metaverse. Section~\ref{sec-MEC-Metaverse-motivation} and~\ref{sec-MEC-Metaverse-SecurityPrivacy} provides the motivation and security/privacy of MEC-enabled Metaverse respectively.

\subsection{Metaverse} \label{sec-Metaverse}

With the rapid development of engine technologies, the ``user-centred” Metaverse has received considerable attention among the commercial and academic communities. Based on these core technologies, such as augmented reality, virtual reality, digital twin, blockchain, etc., users can perform the same daily activities as in the real world and build their own virtual social networks in this digital world. For example, users can work in the virtual world as avatars and hold virtual meetings. Digital currencies, NFTs and blockchains have emerged to allow people to trade in the Metaverse. Immersion, materialization, and interoperability are the main characteristics of Metaverse. This emerging virtual system has drastically transformed the way people live their own lifestyles. Typical examples are discussed as follows.

\textbf{Play to Earn.} Recently, play-to-earn (P2E) games have broken away from the inherent weaknesses of conventional online games. In traditional games, players' assets and trading behaviours are regulated by the game company, and there is a huge risk in trading activities between players. The second point is that the value of players' currencies and goods in the game is given by the company, and players do not have the autonomy to control the market. Conversely, emerging blockchain-based metaverse P2E games enable players to purchase and trade virtual in-game props and commodities safely with real-world currency, independent of the limitations imposed by the game companies. Currently, the most popular P2E games are  \textit{Decentraland} ({https://decentraland.org/}), \textit{Axie Infinity} ({https://axieinfinity.com}), and \textit{The Sandbox} ({https://www.sandbox.game/en/}). Axie Infinity, in particular, where players can trade virtual pets in a digital universe using AXS tokens. 

\textbf{Web3 and Web3.0.} Web3, featuring blockchain as the core technology, is an emerging decentralised network. In Web3, ownership is not centralised and users are the centre of the network. Based on token-based economics concepts, this network supports an autonomous economy and safeguards the ownership of users' assets from companies. Currently, multiple operating systems coexist, including Ethereum, Solana and Polygon, and the supported currencies vary between the different operating systems. The existence of multiple systems further enhances the interoperability and generality of the metaverse system. Unlike the concept of Web3, Web3.0, proposed by Tim Berners-Lee, is a generational product of the evolution of the Web, enabling the sharing of data between different applications and communities. Though there are differences between Web3 and Web3.0, their ultimate goal is to improve the Internet.

\subsection{The Need of Mobile Edge Computing for the Metaverse} \label{sec-MEC-Metaverse-motivation}

Conventional computing paradigms have colossal computing resources and are capable of handling intensive computing tasks. However, data communication delays between the clients and the servers are too excessive and can even cause network congestion when passing through the central network. All these drawbacks contribute to the user's perception of the experience in the metaverse. Excessive latency, for example, over 30ms in AR applications and over 20ms in VR applications\cite{mangiante2017vr}, can cause dizziness and nausea for the users.

MEC gets the spotlight to balance the trade-off among computational delay, transmission latency, and power consumption. As mentioned above, the transmission between the MEC servers and the end devices can be executed within milliseconds and with a specific latency. Furthermore, the higher downlink (DL) rate (200Mbps), supported by 5G technology, facilitates a significant increase in the resolution of the rendering virtual environment. The application of mobile edge computing has substantially improved the performance of the metaverse, further enhancing the users' immersion in this system.

\subsection{Security and Privacy in MEC-enabled Metaverse} \label{sec-MEC-Metaverse-SecurityPrivacy}

Security and privacy are two main concerns of the metaverse \cite{wang2022survey}. In this subsection, the details are discussed as follows.

\textbf{Security.}  In the real world, criminals are punished by the appropriate laws. But within the metaverse, attackers receive little or even no punishment. This means that users living in the metaverse are often subject to a wide range of crimes, including harassment, humiliation, rape, bullying and attacks by multiple coordinated users\cite{ling2021first}\cite{weber2021amplifying}. In addition to criminality, the security of users' devices is also critical. Hackers will use the players' headsets to record and film the users' real environment and even alter what the users hear and see to threaten them\cite{kurtunluouglu2022security}. Therefore, securing users and devices against potential attackers is an area of concern.

\textbf{Privacy.} The collection of multi-sensor information increases the users' immersion in the metaverse, but it also raises the risk of leakage of private information about users. Among the privacy issues experienced by users in the metaverse are privacy leakage, digital assets theft, and identity forgery. To provide a more realistic interactive environment, platforms and companies collect personal information about users, including pupils, fingerprints, faces, movement paths\cite{shang2020arspy}, etc. This vast amount of private data is subject to potential attacks during transmission\cite{wei2020ldp} and processing, leading to the leakage of users' private data. Leakage of users' data will lead to harassment with malicious messages, loss of digital assets, and identity theft in the virtual environment. Moreover, users' location and identity information in the real world may be monitored. In addition, users can be affected by malicious data injection and virtual identity forgery in the metaverse, leading to a disturbed virtual life.

\begin{figure}[t]
    \centering
    \includegraphics[width=8cm, height= 5cm]{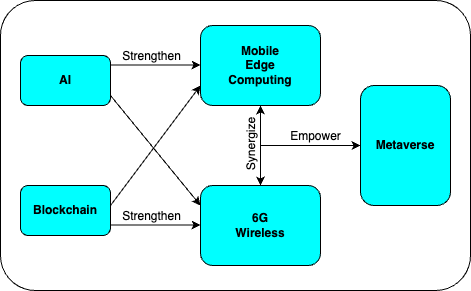}
    \caption{The relationship of AI, Blockchain, MEC, 6G Wireless and Metaverse.}
    \label{fig1}
\end{figure}

\section{Convergence}\label{sectionVI}

This section is devoted to discussing the convergence of the emerging technologies. 

\subsection{6G Wireless Communications with Mobile Edge Computing}

Superpowered traffic applications involve AI-based intelligent decision-making, cellular network-based, shop-floor communication for information interaction and the collection and processing of traffic information, vehicle information and environmental information, which can significantly improve the traffic environment and reduce traffic accidents, by analyzing and predicting traffic conditions as well as a range of other traffic tasks. Based on 6G networks, smart devices can be flexibly networked to use edge computing, AI, blockchain and other technologies for data monitoring and interaction, as well as offloading decisions, enabling self-organization of production and reducing human intervention.

Compared to 5G, 6G performance indicators will be significantly improved, with most performance indicators increasing by 10 to 100 times compared to 5G, such as peak transmission speeds of 100 Gbit/s to 1 Tbit/s for 6G, compared to 10 Gbit/s for 5G. 6G will be highly reliable, with an outage probability of less than one in a million, and positioning-related applications will become more prevalent in 6G, with a corresponding increase in positioning accuracy of a few centimeters indoors and 1 meter outdoors, a tenfold increase compared to 5G. 6G will also support ultra-high density connections, with a density of over 100 devices per cubic metre. In addition, 6G will use the terahertz band for communication and will increase network capacity significantly.

The requirements of 6G applications are more stringent, so it is worthwhile for researchers to investigate the future trend of applying MEC technology to 6G scenarios to provide low-latency, highly reliable services to users.

\subsection{Mobile Edge Computing Strengthened by Artificial Intelligence}
Wu \textit{et al.} \cite{wu2021machine} introduce a novel machine learning prediction technology in mobility management. Machine learning techniques can solve the service discontinuity problem of mobile devices by predicting the uncertainty and the reference signal received power (RSRP). In this paper, neural networks are used for prediction to obtain different RSPP averages and standard deviations for each unit, and the results are analyzed to determine the MEC servers that will be involved and to allow cached content to be sent to the MEC servers in advance in order to reduce the probability of disruption.

In future research directions, mobility management can be linked to artificial intelligence. Through artificial intelligence techniques, the system can learn the users' behaviours on its own, thus predicting how the user will move and transmitting information about the user's state to the SDN controller in advance, thus providing a more reliable service.

\subsection{Mobile Edge Computing Strengthened by Blockchain}

The lack of computation and energy resources of MEC servers causes a decrease in the system efficiency. With the support of blockchain technology, the security of the mobile edge computing paradigm, computing resources utilization and energy consumption efficiency are greatly enhanced. A recent paper \cite{nguyen2022intelligent} presents a novel consensus mechanism based on blockchain mining to enhance the system's security. In addition, the work~\cite{wang2022blockchain} proposes a novel system for addressing the resource allocation issue in polynomial time. Therefore, in the vision for the future, blockchain-based technology to tackle the challenge of limited resources in mobile edge computing is the focus of research.

\subsection{Convergence of the Technologies: AI and Blockchain Assisted Mobile Edge Computing for the Metaverse over 6G Wireless Communications}

Note that this paper covers five technologies: AI, blockchain, mobile edge computing, Metaverse, and 6G wireless communications. In the previous sections, we have elaborated on mobile edge computing,  how AI and blockchain can strengthen it, and how it can be applied to the Metaverse and 6G wireless communications.
In this section, we envision their convergence to have AI and blockchain-aided mobile edge computing for the Metaverse over 6G wireless communications. In such a scenario, the relationships between the technologies are illustrated in Figure~\ref{fig1}. AI and blockchain are both significant technologies to strengthen MEC and 6G wireless networks. The synthesis of MEC and 6G greatly empowers the Metaverse to provide an immersive and high-fidelity performance.

 A recent technical paper~\cite{Chua2022} uses AI for the metaverse over 6G networks. It will be of interest to investigate how blockchain and mobile edge computing can help improve the system in~\cite{Chua2022}. A very recent survey~\cite{chang20226g} presents 6G-enabled edge AI for the Metaverse, but blockchain is not highlighted. In addition,~\cite{tang2022roadmap} focuses on 6G for the Metaverse;~\cite{khan2022metaverse} presents the Metaverse for wireless systems; and~\cite{xu2022full} explains edge-enabled Metaverse. Meanwhile, our work also considers the support of AI and blockchain for MEC in the context of 6G-enabled Metaverse.

    \linespread{.98}

\section{Conclusion}\label{sectionVII}
This paper introduces the background of MEC and compares different computing paradigms over the past decade, and summarizes MEC advantages and disadvantages. Further, we illustrate the  potential of MEC along with AI and blockchain for developing the Metaverse over 6G communications.

As the technologies continue to evolve, mobile networks will usher in a range of exciting new ways to work and play, such as high-fidelity Metaverse experiences based on blockchain and AI technologies, immersive gaming with AR/VR/XR based on 6G wireless networks, as well as traffic assistance systems and smart driving systems integrated with the Internet of Vehicles and machine learning technology, where MEC will also shine. At the same time, new industry standards for MEC and the deployment of MEC platforms will also provide a network ecosystem and value chain, bringing new operating models to major mobile operators, device providers and third-party companies. In the future, MEC will be an integral part of a large platform that brings together computing and communications technologies and will open up endless possibilities for innovation in network services through  integration with other vital technologies.\vspace{0pt}

\section*{Acknowledgement\vspace{0pt}}

This research is supported in part by Nanyang Technological University Startup Grant; and in part by the Singapore Ministry of Education Academic Research Fund under Grants Tier 1 RG97/20,  Tier 1 RG24/20 and  Tier 2 MOE2019-T2-1-176.\vspace{0pt}

\bibliographystyle{IEEEtran}
\bibliography{mybibfile2} 

\end{document}